\documentclass[useAMS,usenatbib,usegraphicx]{mn2e}

\title[Cool white dwarf atmospheres]{The polluted atmospheres of cool white dwarfs and the magnetic field connection\thanks
{Based on observations collected at the European Organisation for
Astronomical Research in the Southern Hemisphere, Chile under programmes
ID 082.D-0750 and 091.D-0267.}}
\author[A. Kawka and S. Vennes]{Ad\'ela Kawka$^{1}$\thanks{E-mail:
kawka@sunstel.asu.cas.cz (AK); vennes@sunstel.asu .cas.cz (SV)} and St\'ephane
Vennes$^{1}$\footnotemark[2] \\
$^{1}$Astronomick\'y \'ustav, Akademie v\v{e}d \v{C}esk\'e republiky, Fri\v{c}ova 298, CZ-251 65 Ond\v{r}ejov, Czech Republic }
\begin{document}

\date{}

\pagerange{\pageref{firstpage}--\pageref{lastpage}} \pubyear{2013}

\maketitle

\label{firstpage}

\begin{abstract}
We present an analysis of X-Shooter spectra of the polluted hydrogen-rich white 
dwarfs (DAZ) NLTT~888 and NLTT~53908. The spectra of NLTT~53908 show strong, Zeeman-split calcium lines (Ca\,{\sc ii}~H\&K and 
Ca\,{\sc i} $\lambda$4226) and the star appears to be a close relative of the polluted magnetic white dwarf (DAZH) NLTT~10480, 
while the spectra of NLTT~888 show narrow lines of calcium and iron. 
A comparison of the DAZ NLTT~888 and the DAZH NLTT~53908 with other class members
illustrates the diversity of environment and formation circumstances surrounding these objects. 
In particular, we find that the incidence of magnetism in old, polluted white dwarfs significantly exceeds 
that found in the general white dwarf population which suggests an hypothetical link between a crowded planetary 
system and magnetic field generation.
\end{abstract}


\section{Introduction}

Polluted white dwarfs provide an opportunity to investigate the ultimate fate of
planetary systems. \citet{vil2007} show that under the right conditions, some
planets do manage to survive the asymptotic giant branch (AGB) and post-AGB
phases, while \citet{deb2002} had already shown that these planets and asteroids could find
themselves within the white dwarf tidal radius if their orbits become
unstable during close encounters with each other. The formation of debris discs around white dwarfs
may be a common occurrence \citep{jur2003,kil2006} and is most likely related to the presence of a crowded circumstellar environment in the parent star.
The presence of debris material accreted onto the white dwarf surface is evident in spectroscopic observations
of a large fraction of hydrogen-rich white dwarfs \citep{zuc2003}. 

The origin of magnetic fields in white dwarf stars may
be linked to possible field-generating merger events preceding the birth of the white dwarf
\citep{tou2008,nor2011,gar2012}.  The likelihood of such mergers
during the post-AGB phase would be directly related to the multiplicity in their respective stellar or planetary systems:
Surface metallicity and magnetic field in white dwarfs may be concomitant.

Recent investigations have uncovered several cool, magnetic, polluted DA white dwarfs
\citep{far2011,kaw2011,zuc2011}. Because the initial class identifications were secured with low-dispersion
spectra, the present
DAZ selection is independent of field intensity weaker than $\sim 1$ MG. 
The inverse problem remains to be investigated: Could some high-field white dwarfs also be polluted? An answer
to that question would require detailed modeling beyond linear Zeeman effect, but spectral lines 
could still escape detection because of line spread. 

We present an analysis of two recently identified polluted white dwarfs from the 
revised NLTT catalogue of \citet{sal2003}:  NLTT~888 and the magnetic
white dwarf NLTT~53908. 
In Section 2, we present new high signal-to-noise ratio spectra of these objects. With these data, we constrain the atmospheric 
parameters (Section 3.1), the magnetic field strength in NLTT~53908 (Section 3.2), and we conduct
an abundance analysis (Section 3.3). Finally we summarize
our results in Section 4 and examine the particular case of NLTT~53908 along with the class properties of cool, magnetic 
DAZ white dwarfs.

\section{Observations}

\begin{figure*}
\includegraphics[viewport=0 200 570 535,clip,width=0.79\textwidth]{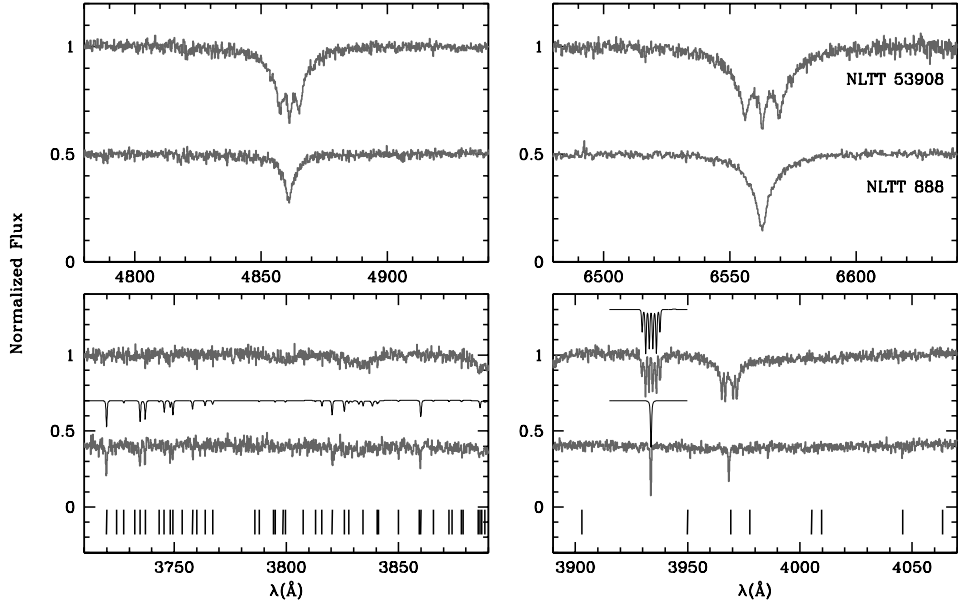}
\vspace{-0.2cm}
\caption{Sections of the X-shooter spectra (grey lines) of NLTT~888 (offset $-0.5$ in the top panels and $-0.6$ in the lower panels) and NLTT~53908 from 
3710 to 6650 \AA. All iron lines with a strength $\log{gf} < -1.2$ and a lower energy level $E_{\rm low} < 20,000$ cm$^{-1}$ are marked with
vertical lines. Table~\ref{tbl_lines} lists all prominent lines. The lower panels also show best fitting models to calcium and iron lines (black lines) shifted by $+0.3$ relative to the observed spectra (see Section 3.3).
\label{fig1}}
\end{figure*}

We first observed NLTT~53908 with the ESO Faint Object Spectrograph and Camera
(EFOSC2) attached to the New Technology Telescope (NTT) at La Silla 
Observatory on UT 21 October 2008. We used grism number 11 (300 lines/mm) with 
a 1 arcsecond slit-width resulting in a resolution of $\Delta \lambda \sim 14$ \AA. 
The spectrum revealed a cool DAZ white dwarf with Ca\,{\sc ii} H\&K.
Incidentally, \citet{lim2013} classified NLTT~53908 (PM I 22276$+$1753) as a DA white dwarf.

\begin{table}
\centering
\begin{minipage}{\columnwidth}
\caption{Equivalent widths and line velocities of NLTT~888.\label{tbl_lines}}
\begin{tabular}{@{}l@{\hskip 0.3cm}c@{\hskip 0.3cm}c@{\hskip 0.3cm}l@{\hskip 0.3cm}c@{\hskip 0.3cm}c@{}}
\hline
Ion/$\lambda$$^a$ & E.W.  & $v$  & Ion/$\lambda$$^a$ & E.W.  & $v$ \\
 (\AA)  & (m\AA)      &  (km\,s$^{-1}$) &  (\AA)  &  (m\AA)  &  (km\,s$^{-1}$) \\
\hline
Fe\,{\sc i}\,3440.735$^b$& 144. & 18.7 & Fe\,{\sc i}\,3820.425 &  83. & 26.3 \\
Fe\,{\sc i}\,3581.195& 133. &  9.6 & Fe\,{\sc i}\,3859.911 &  75. & 13.9 \\
Fe\,{\sc i}\,3719.935& 134. & 19.7 & Ca\,{\sc ii}\,3933.66 & 196. & 18.0 \\
Fe\,{\sc i}\,3734.864&  90. & 16.6 & Ca\,{\sc ii}\,3968.47 & 140. & 19.7 \\
Fe\,{\sc i}\,3737.131&  73. & 19.2 & Fe\,{\sc i}\,4045.813 &  43. & 13.9 \\
Fe\,{\sc i}\,3748.262&  58. & 19.8 & Fe\,{\sc i}\,4063.594 &  30. &  7.8 \\
Fe\,{\sc i}\,3749.485&  51. & 14.8 & H\,{\sc i}\,4861.323  &  ... &  7.8 \\
Fe\,{\sc i}\,3758.233&  51. & 26.9 & H\,{\sc i}\,6562.797  &  ... & 24.2 \\
\hline
\end{tabular}\\
$^a$ Laboratory wavelength from the National Institute of Standards and 
Technology (NIST) which is accessible from http://www.nist.gov/pml/data/asd.cfm.\\
$^b$ Blend of Fe\,{\sc i} 3440.606 and 3440.989 \AA.
\end{minipage}
\end{table}

Following-up on our own classification, we obtained two sets of echelle spectra of NLTT~53908 using the X-shooter 
spectrograph \citep{ver2011} attached to the UT2 (Kueyen) at Paranal 
Observatory on UT 10 July 2013 and 9 August 2013. The spectra were obtained 
with the slit-width set to 0.5, 0.9 and 0.6 arcseconds for the UVB, VIS, and 
NIR arms, respectively. This setup delivered a resolution of 
$R= \lambda/\Delta\lambda = 9900$, 7450 and 7780 for the UVB, VIS and NIR arms, respectively. 
The exposure times were 2940 and 3000 s for the UVB and VIS arms, respectively.
For the NIR arm we obtained five exposures of 600 s each. The 
observations were conducted at the parallactic angle.

\begin{table}
\centering
\begin{minipage}{\columnwidth}
\caption{Photometry\label{tbl_phot}}
\begin{tabular}{@{}lccc@{}}
\hline
Survey/band$^a$ & $\lambda$ eff. & \multicolumn{2}{c}{Magnitude} \\
   &  & NLTT~888 & NLTT~53908 \\
\hline
{\it GALEX} NUV & 2315 \AA     & $22.416\pm0.264$ & $19.459\pm0.077$ \\
SDSS $u$        & 3551 \AA     & $19.011\pm0.028$ & $17.522\pm0.011$ \\
SDSS $g$        & 4686 \AA     & $17.983\pm0.006$ & $16.897\pm0.004$ \\
SDSS $r$        & 6166 \AA     & $17.536\pm0.006$ & $16.659\pm0.005$ \\
SDSS $i$        & 7481 \AA     & $17.352\pm0.006$ & $16.590\pm0.005$ \\
SDSS $z$        & 8931 \AA     & $17.344\pm0.017$ & $16.612\pm0.010$ \\
2MASS $J$       & 1.235 $\mu$m & $16.478\pm0.104$ & $15.853\pm0.076$ \\
2MASS $H$       & 1.662 $\mu$m & $16.348\pm0.206$ & $15.567\pm0.117$ \\
2MASS $K$       & 2.159 $\mu$m &  17.058:         & $15.466\pm0.189$ \\
{\it WISE} $W1$ & 3.353 $\mu$m & $16.293\pm0.091$ & $15.088\pm0.045$ \\
{\it WISE} $W2$ & 4.603 $\mu$m & $16.032\pm0.254$ & $15.313\pm0.121$ \\
\hline
\end{tabular}\\
$^a$ {\it GALEX} GR6/GR7 photometry obtained at 
galex.stsci.edu/GalexView/; SDSS Photometric Catalog, Release 10 \citep{ahn2013};
2MASS photometry \citep{skr2006}; ({\it WISE}) photometry \citep{cut2012}. 
\end{minipage}
\end{table}

We first observed NLTT~888 with FORS2 attached to the UT2. Based on these
spectropolarimetric data, \citet{kaw2012a} placed an upper limit of 40 kG on the 
longitudinal magnetic field. The blue FORS2 spectrum revealed a cool 
DAZ white dwarf showing Ca\,{\sc ii} H\&K absorption lines with an
abundance of $\log{\rm Ca/H} = -10.65\pm0.15$ \citep{kaw2012a}. Consequently,
we obtained a set of six X-shooter spectra on UT 4 August 2013,
2 September 2013 and 3 September 2013. 
The observations were conducted with the configuration and exposure times described earlier.
The spectra (Fig.~\ref{fig1}) confirmed the Ca\,{\sc ii}
line identifications and, in addition, revealed several Fe\,{\sc i} lines. 
Table~\ref{tbl_lines} lists all strong lines along with their respective equivalent widths and
heliocentric velocities.
The H$\alpha$ core appears rounded or possibly split by the effect of a $\sim$30 kG field.
The heavy element line velocities average $17.5\pm5.2$\,km\,s$^{-1}$ and are
marginally consistent with the H$\alpha$ velocity ($23.4\pm2.0$ \,km\,s$^{-1}$).

Table~\ref{tbl_phot} lists available photometric measurements from the Galaxy 
Evolution Explorer ({\it GALEX}) sky survey, the Sloan Digital Sky Survey 
(SDSS), the Two Micron All Sky Survey (2MASS), and the Wide-field Infrared 
Survey Explorer ({\it WISE}). We describe in Section 3.1 the continuum temperature 
measurements using the photometric data.

\section{Analysis}

In our analysis of the atmospheric properties of NLTT~888 and NLTT~53908
we used a grid of hydrogen-rich model atmospheres calculated in local 
thermodynamic equilibrium. These models are described in \citet{kaw2006} and 
\citet{kaw2012b}.
We calculated masses and cooling ages using the evolutionary mass-radius 
relations of \citet{ben1999}.

\subsection{Atmospheric Parameters and Spectral Energy Distributions (SED)}

\begin{figure}
\includegraphics[width=1.0\columnwidth]{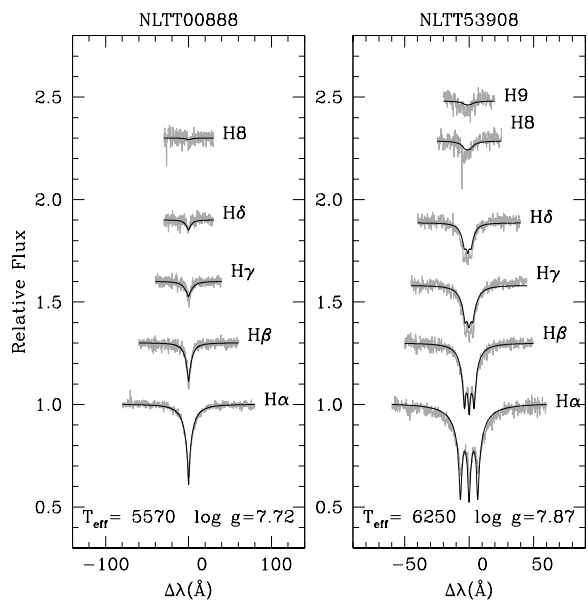}
\vspace*{-0.7cm}
\caption{Balmer lines of NLTT~888 (left) and NLTT~53908 (right) compared
to their corresponding best-fit model spectra.
\label{fig2}}
\end{figure}

First, we determined the effective temperature and surface gravity of NLTT~888 and
NLTT~53908 by fitting the observed Balmer lines with grids of synthetic spectra
using $\chi^2$ minimization techniques (Fig.~\ref{fig2}). For NLTT~888 we included H$\alpha$
to H8, but excluded H$\epsilon$ which is dominated by Ca\,{\sc ii} K. For
NLTT~53908 we included H$\alpha$ to H9, but again excluded H$\epsilon$.
For NLTT~53908 we also included the effect of a magnetic field on the line
formation as described by \citet{unn1956} and \citet{mar1981}.
However, the H$\alpha$ core is poorly fitted:
We neglected the effect of a field distribution associated with, for example,
a dipolar field structure that would broaden the observed line profile.
Instead, we assumed a single valued field. 

\begin{figure}
\includegraphics[width=0.94\columnwidth]{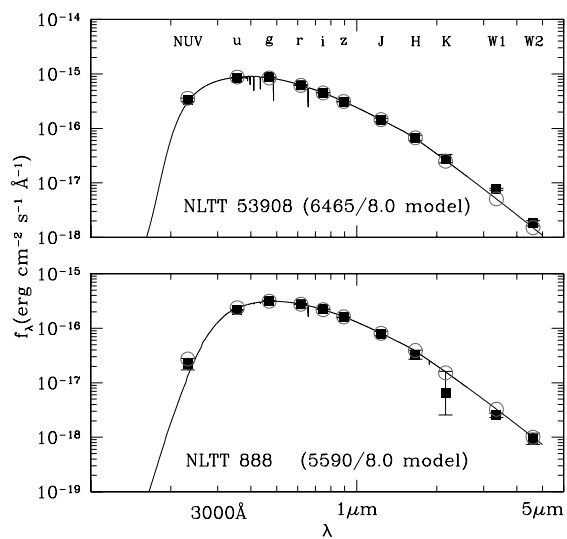}
\vspace{-0.5cm}
\caption{SED, $f_\lambda$ (erg cm$^{-2}$ s$^{-1}$ \AA$^{-1}$) vs $\lambda$,
of NLTT~888 (bottom) and NLTT~53908 (top).  The observed SED (full squares) 
was built using photometric data (Table~\ref{tbl_phot}), and the synthetic 
SED (open circles) was computed using best-fit model (full line). We noted
the possible detection of a mild infrared excess in NLTT~53908.
\label{fig3}}
\end{figure}

Figure~\ref{fig3} shows the observed and calculated SED for both objects.
In the case of NLTT~888, the effective temperature determined from the Balmer line fits 
($5570$ K) is consistent with the temperature determined from the SED 
($5590$ K). However, for NLTT~53908, the effective temperature from the 
Balmer line fits ($6250$ K) is lower than the effective
temperature required to match the observed SED (6465~K).
Table~\ref{tbl_prop} summarizes these measurements and lists additional properties
derived from these measurements.

\subsection{Magnetic field}

We measured the strength of the average surface magnetic field in NLTT~53908 
using the Zeeman split pattern of H$\alpha$ and the calcium lines. Details of 
the predicted Zeeman line components are presented in \citet{kaw2011}.
We fitted Gaussian functions to each Zeeman component of Ca\,{\sc i} $\lambda$ 4226 \AA,
Ca\,{\sc ii} H\&K, and H$\alpha$, and measured the line centres. Next, we simultaneously constrained
the magnetic field strength and average line velocity by fitting the observed line centres
with predicted positions.
Using the calcium lines we determined a surface
averaged magnetic field of $B_S=0.335\pm0.003$ MG with a velocity of 
$v=19.8\pm1.7$ km~s$^{-1}$. Using H$\alpha$ we determined 
$B_S=0.331\pm0.004$ MG with a velocity of $v=23.9\pm2.9$ km~s$^{-1}$. The 
magnetic field measurements and velocities are consistent within uncertainties.
We adopted the weighted average 
of these measurements $B_S=0.334\pm0.003$ MG and $v=20.8\pm1.5$ km~s$^{-1}$.

The H$\alpha$ core in NLTT~888 shows the possible effect of broadening due to 
unresolved splitting by a weak field
($\sim30$ kG). However, a higher resolution spectrum would be required to fully
resolve the Zeeman pattern and confirm the magnetic field detection.

\subsection{Abundance analysis}

We initiated a study of the abundance patterns in NLTT~888 and 53908 by fitting a
set of model spectra with varying abundances to the X-shooter spectra. For 
NLTT~888 we set the effective temperature and surface gravity to that 
determined from the Balmer line fits. However, for NLTT~53908 we adopted 
$T_{\rm eff}=6300$~K ($\log{g}=7.8$) as a compromise between the Balmer line fit and the SED analysis. 
For reference, we adopted the solar abundance scale described in \citet{ven2013}
that was built using the compilations of \citet{asp2009} and \citet{lod2009}.
The measured abundances and upper limits are summarized in Table~\ref{tbl_prop}.
The quoted uncertainties are statistical only (1$\sigma$), and we may
add systematic errors of $\sim0.1$ dex due to model uncertainties (see below).
The upper limits are taken at the 90\% confidence level.

The measured calcium ($\log{\rm Ca/H} = -10.77$) and iron ($\log{\rm Fe/H} = -9.01$) abundances 
imply that the atmosphere of NLTT~888 is iron enhanced ($\log{\rm Fe/Ca} = 1.76$). Note that if we assume 
$\log{g} = 8$ the abundances would be shifted up by 0.1 dex, while abundance 
ratios remain unaffected. The calcium abundance measured with X-shooter 
is in agreement with that
originally determined by \citet{kaw2012a} using FORS data ($\log{\rm Ca/H} = -10.65\pm0.15$).
We calculated magnesium and silicon abundance upper limits, but were unable
to secure meaningful constraints to the aluminium abundance. 

Next, we measured the abundance of calcium in the atmosphere of NLTT~53908 and calculated upper limits
for the abundance of magnesium, silicon and iron. We found a significant discrepancy in the
abundance of calcium based on Ca\,{\sc i} and Ca\,{\sc ii} lines. Using the 
Ca\,{\sc ii} H\&K lines
we measured $\log{\rm Ca/H} = -9.85\pm0.04$ (${\rm [Ca/H]} = -4.18\pm0.04$) compared to 
$\log{\rm Ca/H} = -9.24^{+0.10}_{-0.19}$ 
(${\rm [Ca/H]} = -3.57^{+0.10}_{-0.19}$) using the Ca\,{\sc i}$\lambda 4226$ line. A similar problem
was also encountered in our analysis of the cool magnetic DAZ NLTT~10480 \citep{kaw2011}. In this case, adopting 
a cooler atmosphere helped restore the calcium ionization balance. However, in the case
of NLTT~53908 a lower effective temperature that would restore consistency is excluded by the SED and Balmer line fits.
A surface gravity shift of 0.3 dex results in abundance shift of 0.06 dex in both Ca\,{\sc i}
and Ca\,{\sc ii} measurements, leaving the ionization ratio virtually unchanged. Also, a change
in temperature of $-200$~K results in abundance shifts of $0.25$ and $0.15$ using Ca\,{\sc i} and Ca\,{\sc ii} lines, respectively, corresponding to a change in the Ca\,{\sc i}$/$Ca\,{\sc ii} ionization ratio of 0.1 dex. Such small variations
do not resolve the present difficulties with the calcium ionization balance.
Because Ca\,{\sc ii} is the dominant ionization species
we adopted the abundance measurement based on the Ca\,{\sc ii} lines.
The calcium abundance and the upper limits offer a glimpse of the
composition of NLTT~53908 and we may
conclude that its atmosphere is not iron enhanced.

\begin{table}
\centering
\begin{minipage}{\columnwidth}
\caption{Properties \label{tbl_prop}}
\begin{tabular}{@{}llc@{}}
\hline
Parameter & NLTT~888 & NLTT~53908 \\
\hline
$T_{\rm eff}$ (K)  & $5570\pm40$    & $6250\pm70$     \\
$\log{g}$ (cgs)    & $7.72\pm0.08$  & $7.87\pm0.12$   \\
Mass ($M_\odot$)   & $0.45\pm0.03$  & $0.51\pm0.07$   \\
Distance (pc)      & $58\pm3$       & $38\pm3$        \\
Age (Gyr)          & $4.0^{+1.8}_{-0.3}$ & $1.7^{+0.5}_{-0.3}$ \\
$T_{\rm eff}$ (K)$^a$ & $5590\pm50$ & $6465\pm55$  \\
$\log{g}$ (cgs)$^a$   & (8.0)       & (8.0)       \\
$\log{\rm Mg/H}$ (${\rm [Mg/H]}^b$) & $< -8.7(-4.3)$ & $< -7.9(-3.5)$ \\
$\log{\rm Si/H}$ (${\rm [Si/H]}^b$) & $< -8.9(-4.4)$ & $< -8.0(-3.5)$ \\
$\log{\rm Ca/H}$ (${\rm [Ca/H]}^b$) & $-$10.77($-$5.10)$\pm$0.06 & $-$9.85($-$4.18)$\pm$0.04 \\
$\log{\rm Fe/H}$ (${\rm [Fe/H]}^b$) & $-$9.01($-$4.48)$\pm$0.07 & $< -8.7(-4.2)$ \\
$v_r$ (km\,s$^{-1}$) & $23.4\pm2.0$ & $20.8\pm1.5$      \\  
$B_s$ (kG)         &    $<40$       & $334\pm3$        \\ 
\hline
\end{tabular}\\
$^a$ Based on the SED. \\
$^b$ [X/H] $= \log{\rm X/H} - \log{\rm X/H}_\odot$.
\end{minipage}
\end{table}

\begin{table*}
\centering
\begin{minipage}{\textwidth}
\caption{Known cool DAZ white dwarfs ($T_{\rm eff} < 7000$ K). \label{tbl_cool}}
\begin{tabular}{@{}llccclccc@{}}
\hline
WD & Name & $T_{\rm eff}$ (K) & $\log{g}$ (c.g.s) & $\log{\rm (Ca/H)}$ & Fe$/$Ca & $B_s$ (kG) & $|B_l|$ (kG) &  Reference \\
\hline
WD0015$-$055 & NLTT~00888 & $5570\pm40$  & $7.72\pm0.08$ & $-10.77\pm 0.06$  & 58    &  $<40$     & $<40$       &  1,2 \\
WD0028$+$220 & NLTT~01675 & $6020\pm50$  & $8.04\pm0.07$ & $-9.53\pm0.03$  &  8    &  $<40$     & ...         &  3 \\
WD0151$-$308 & NLTT~06390 & $6040\pm40$  & $7.90\pm0.07$ & $-10.00\pm0.04$ & 27    &  $<40$     & $<11$       &  2,3 \\
WD0243$-$026 & LHS~1442   & $6800\pm300$ & $8.15\pm0.10$ & $-9.90$  & ...   &  $<10$     & ...         &  4 \\
WD0245$+$541 & G~174-14   & $5190\pm300$ & $8.22\pm0.10$ & $-12.69$  & ...   &  $<10$     & ...         &  4 \\
WD0315$-$293 & NLTT~10480 & $5200\pm200$ & $8.0\pm0.5$   & $-10.3\pm0.3$   & $<$10 &  $519\pm4$ & $212\pm50$  &  3,5 \\
WD0322$-$019 & G~77-50    & $5310\pm100$ & $8.05\pm0.01$ & $-9.8\pm0.2$    & 13    &  120       & ...         &  6 \\
WD0334$-$224 & NLTT~11393 & $5890\pm30$  & $7.86\pm0.06$ & $-10.24\pm0.04$ & $<$7  &  $<40$     & $<16$       &  2,3 \\
WD1208$+$576 & G~197-47   & $5830\pm300$ & $7.91\pm0.10$ & $-10.96$  & ...   &  $<10$     & ...         &  4 \\
WD1344$+$106 & G~63-54    & $6945\pm300$ & $7.99\pm0.10$ & $-11.13$  & ...   &  $<10$     & ...         &  4 \\
WD1633$+$433 & G~180-63   & $6570\pm300$ & $8.08\pm0.10$ & $-8.63$   & 5     &  $<10$     & ...         &  4 \\
WD1653$+$385 & NLTT~43906 & 5900         & 8.0           & $-7.9\pm0.19$   & 1.3   &   70       & ...         &  7 \\
WD2225$+$176 & NLTT~53908 & $6250\pm70$  & $7.87\pm0.12$ & $-9.85\pm 0.04$   & $<$13 &  $334\pm3$ & ...         &  1 \\
\hline
\end{tabular}\\
References: (1) This work; (2) \citet{kaw2012a}; (3) \citet{kaw2012b}; (4) \citet{zuc2003}; (5) \citet{kaw2011}; (6) \citet{far2011};
(7) \citet{zuc2011}
\end{minipage}
\end{table*}

\section{Summary and Discussion}

We obtained accurate calcium abundance measurements in the atmospheres of the 
DAZ white dwarfs NLTT~53908 and NLTT~888. Also, we measured the iron abundance 
in NLTT~888 and found that NLTT~53908 is a magnetic white dwarf. These two 
objects join a class of cool, polluted white dwarfs. Table~\ref{tbl_cool} lists 
all DAZ white dwarfs with $T_{\rm eff} < 7000$ K. Their atmospheric parameters 
and abundance measurements are taken from the cited references. It is worth 
noting that within this sample of cool DAZ white dwarfs, the iron to calcium 
abundance ratio varies by a factor of $\sim50$ from an abundance ratio close 
to unity (NLTT~43906) and up to $\sim$60 (NLTT~888),
and assuming steady state accretion, this range of abundances
suggests a remarkable diversity in the accretion source composition.

Also, Table~\ref{tbl_cool} lists confirmed surface magnetic fields
$B_s$ or upper limits based on the shape 
of the H$\alpha$ core. Requiring that the $\sigma$ components occupy 
separate resolution elements we have $\lambda R^{-1} = \Delta \lambda_B$, 
where $R$ is the resolving power and $\lambda_B$ is the magnetic line 
displacement. We find that $B\la 235\,g/\lambda$\,kG, where $g$ is the
Land\'e factor and $\lambda$ is given 
in \AA. For those stars that have been observed spectropolarimetrically we 
provide the longitudinal magnetic field $|B_l|$ measurement or upper limit. 
Out of 13 objects listed in Table~\ref{tbl_cool}, 
four are known to possess a 
magnetic field. Restricting the sample to temperatures below 6000 K, 
seven objects remain with three harbouring a magnetic
field. 

The incidence of magnetism in the general population of white dwarfs 
has been reported to be as low as $\sim5$\% in spectropolarimetric surveys \citep{sch1995,kaw2012a} to possibly as high as 
13-21\% in the old population of nearby white dwarfs \citep{kaw2007}. 
The field strengths range from a few kG \citep[see, e.g., ][]{lan2012} to 
several 100 MG covering five orders of magnitude. 
Assuming a flat distribution per 
field decade \citep{sch1995,kaw2007}, we estimate that 40\% of magnetic white 
dwarfs have fields below 1 MG (corresponding to the largest field in our DAZ 
sample), i.e., $\sim$7\% of the old, local population. Assuming that our DAZ sample 
was drawn from a similar population we find that the probability of identifying 
4 magnetic white dwarfs in a sample of 13 objects is below 1\%. Therefore, we 
can be confident at the 99\% level that the population of cool DAZ white dwarfs 
has a higher incidence of magnetism than currently reported in the general population. 
For example, \citet{ber1997} reported the identification of twelve magnetic white dwarfs in a
survey of $\sim110$ cool white dwarfs (4,000$\ga T_{\rm eff} \ga$10,000 K) or an incidence of
11\%. More specifically, eight out of 64 hydrogen-rich white dwarfs (13\%) were found to be magnetic,
with only two objects with a field below 1~MG. The probability that 
our sample of 4 magnetic objects out of 13 DAZ white dwarfs could be drawn from a population with a 3\% (2/64)
field incidence is below 0.1\%. Important questions remain unanswered: First, having estimated the field incidence
among cool, polluted white dwarfs, what would be the incidence among their warmer ($T_{\rm eff} \la 10,000$~K) counterparts? 
Conversely, what would be the fraction of polluted white dwarfs in a high-field ($>1$~MG) sample?
In the context of merger-induced magnetic fields \citep{tou2008,nor2011,gar2012,kul2013,wic2014}, only mergers occurring on the red giant or
asymptotic red giant phases generate fields. Also, diffusion models \citep[see, e.g., ][]{koe2009} imply that
atmospheric pollution quickly recedes after an accretion event. A dense circumstellar environment 
increases the probability that these events would be observed simultaneously in a given system.

\section*{Acknowledgments}

A.K. and S.V. are supported by the GA\v{C}R
(P209/12/0217, 13-14581S) and the project RVO:67985815.

This publication makes use of data products from {\it WISE}, which
is a joint project of the University of California, Los Angeles, and the Jet Propulsion
Laboratory/California Institute of Technology, funded by the National Aeronautics and Space Administration,
and from 2MASS, which is a joint project of
the University of Massachusetts and the Infrared Processing and Analysis Center/California Institute of
Technology, funded by the National Aeronautics and Space Administration and the National Science Foundation.

\label{lastpage}

\end{document}